\documentclass[%
 reprint,
superscriptaddress,
nofootinbib,
nobibnotes,
 amsmath,
 amssymb,
 aps,
 prx,
]{revtex4-2}

\usepackage{csquotes}
\usepackage{siunitx}
\usepackage{IEEEtrantools}
\usepackage{braket}
\usepackage{amsmath}
\usepackage{amssymb}
\usepackage{nicefrac}
\usepackage{verbatim}
\usepackage{amsmath}
\usepackage{wasysym}
\usepackage{upgreek}
\usepackage{xcolor}
\usepackage{braket}
\usepackage[hidelinks]{hyperref}

\usepackage{filecontents}

\usepackage{float}
\usepackage{subfigure}

\usepackage{graphicx}
\usepackage{dcolumn}
\usepackage{bm}

\DeclareSIUnit\gauss{G}
\DeclareSIUnit\cps{cps}

\begin{document}
\title{Long-Lived Quantum Memory Enabling Atom-Photon \\ Entanglement over 101 km Telecom Fiber}

\author{Yiru Zhou}
\altaffiliation{yiru.zhou@physik.uni-muenchen.de}
\affiliation{Fakult{\"a}t f{\"u}r Physik, Ludwig-Maximilians-Universit{\"a}t M{\"u}nchen, Schellingstr. 4, 80799 M{\"u}nchen, Germany}
\affiliation{Munich Center for Quantum Science and Technology (MCQST), Schellingstr. 4, 80799 M{\"u}nchen, Germany}

\author{Pooja Malik}
\affiliation{Fakult{\"a}t f{\"u}r Physik, Ludwig-Maximilians-Universit{\"a}t M{\"u}nchen, Schellingstr. 4, 80799 M{\"u}nchen, Germany}
\affiliation{Munich Center for Quantum Science and Technology (MCQST), Schellingstr. 4, 80799 M{\"u}nchen, Germany}

\author{Florian Fertig}
\affiliation{Fakult{\"a}t f{\"u}r Physik, Ludwig-Maximilians-Universit{\"a}t M{\"u}nchen, Schellingstr. 4, 80799 M{\"u}nchen, Germany}
\affiliation{Munich Center for Quantum Science and Technology (MCQST), Schellingstr. 4, 80799 M{\"u}nchen, Germany}

\author{Matthias Bock}
\affiliation{Fachrichtung Physik, Universit{\"a}t des Saarlandes, Campus E2.6, 66123 Saarbr{\"u}cken, Germany}

\author{Tobias Bauer}
\affiliation{Fachrichtung Physik, Universit{\"a}t des Saarlandes, Campus E2.6, 66123 Saarbr{\"u}cken, Germany}

\author{\\Tim van Leent}
\affiliation{Fakult{\"a}t f{\"u}r Physik, Ludwig-Maximilians-Universit{\"a}t M{\"u}nchen, Schellingstr. 4, 80799 M{\"u}nchen, Germany}
\affiliation{Munich Center for Quantum Science and Technology (MCQST), Schellingstr. 4, 80799 M{\"u}nchen, Germany}

\author{Wei Zhang}
\altaffiliation{changong@ustc.edu.cn}
\affiliation{Fakult{\"a}t f{\"u}r Physik, Ludwig-Maximilians-Universit{\"a}t M{\"u}nchen, Schellingstr. 4, 80799 M{\"u}nchen, Germany}
\affiliation{Munich Center for Quantum Science and Technology (MCQST), Schellingstr. 4, 80799 M{\"u}nchen, Germany}
\affiliation{Present address: Hefei National Laboratory, Hefei 230088, China}

\author{Christoph Becher}
\affiliation{Fachrichtung Physik, Universit{\"a}t des Saarlandes, Campus E2.6, 66123 Saarbr{\"u}cken, Germany}

\author{Harald Weinfurter}
\altaffiliation{h.w@lmu.de}
\affiliation{Fakult{\"a}t f{\"u}r Physik, Ludwig-Maximilians-Universit{\"a}t M{\"u}nchen, Schellingstr. 4, 80799 M{\"u}nchen, Germany}
\affiliation{Munich Center for Quantum Science and Technology (MCQST), Schellingstr. 4, 80799 M{\"u}nchen, Germany}
\affiliation{Max-Planck Institut f{\"u}r Quantenoptik, Hans-Kopfermann-Str. 1, 85748 Garching, Germany}

\date{\today}

\begin{abstract}
Long-distance entanglement distribution is the key task for quantum networks, enabling applications such as secure communication and distributed quantum computing.
Here we report on novel developments extending the reach for sharing entanglement between a single $^{87}$Rb atom and a single photon over long optical fibers.
To maintain a high fidelity during the long flight times through such fibers, the coherence time of the single atom is prolonged to \SI{7}{ms} by applying a long-lived qubit encoding.
In addition, the attenuation in the fibers is minimized by converting the photon's wavelength to the telecom S-Band via polarization-preserving quantum frequency conversion.
This enables to observe entanglement between the atomic quantum memory and the emitted photon after passing 101 km of optical fiber with a fidelity better than 70.8$\pm$2.4\%.
The fidelity, however, is no longer reduced due to loss of coherence of the atom or photon but in the current setup rather due to detector dark counts, showing the suitability of our platform to realize city-to-city scale quantum network links.
\end{abstract}

\maketitle

\section{Introduction}
Future quantum networks will build on the generation of shared entanglement between distant locations~\cite{Wehner:18}, enabling tasks such as distributed quantum computing~\cite{Monroe:14,Cuomo:20}, quantum cryptography~\cite{Acin:07, Nadlinger:21,Zhang:21}, and quantum sensing~\cite{Proctor:18,Khabiboulline:19}. The entanglement is distributed best via photonic channels using free-space satellite connections~\cite{yin2020entanglement} or optical fiber links~\cite{neumann2022continuous}. While attenuation losses limit point-to-point channel lengths, quantum memories and quantum gates at intermediate nodes enable one to overcome the exponential decay in distribution efficiency by enabling quantum repeater schemes~\cite{Briegel:98,Loock2020}.
Neutral atoms have emerged as a highly promising platform to realize a quantum repeater network~\cite{Covey:23} by demonstrating its key ingredients such as efficient light-matter interfaces~\cite{Brekenfeld:20}, long memory-storage times~\cite{barnes2022assembly}, arrays of individually addressable atoms~\cite{bluvstein2022quantum, scholl2021quantum}, and high fidelity gates~\cite{madjarov2020high,Evered:23}.
So far, mainly segments of quantum networks were realized using, e.g., atoms, ions, or solid-state systems --- including a quantum repeater node~\cite{langenfeld2021quantum, Krutyanskiy:22}, the generation of high-fidelity entanglement of two distant memories~\cite{Stephenson:20,Langenfeld:21,krutyanskiy2023entanglement,Liu:21}, also using photons at telecom wavelength~\cite{Lago-Rivera:21} and over tens of kilometers of fiber~\cite{Leent:22,Luo:22}, and the first three-node networks~\cite{Pompili:21,hermans2022qubit}.

Today, fiber-based communication reliably connects distant locations employing telecom wavelengths to minimize losses.
To achieve maximum distances between the nodes in future quantum networks, it will be equally important to operate at these low-loss wavelengths.
Quantum memories and processors at the nodes, therefore, should either operate at these wavelengths~\cite{gritsch2022narrow} or provide quantum frequency conversion (QFC) to telecom wavelengths~\cite{ikuta2018polarization,Bock:18}.
Recent developments have already enabled memory-photon entanglement distribution over tens of kilometers of optical fiber~\cite{Krutyanskiy:19,Yu:20,Leent:20}. Note, since losses in the quantum channel are inevitable, a classical signal heralding successful entanglement distribution has to be provided by the system to trigger further processing. In addition, the larger the network, the longer is the communication times for such a signal, making long coherence times of the memory a must. For example, assuming a speed-of-light in an optical fiber of $\frac{2}{3}c$, bridging a fiber length on the order of \SI{100}{\kilo\meter} takes around \SI{500}{\micro\second}. Also, in view of providing the entanglement in a quantum network for a certain storage time, it is essential to maintain coherence even much longer than this.

Here we demonstrate a single-atom based quantum network node that allows to distribute entanglement over telecom fiber links with a length of up to \SI{101}{\kilo\meter}. To achieve this, we develop a long-lived qubit encoding to increase the coherence time for the atomic quantum memory~\cite{Koerber:18} and a highly efficient quantum-state preserving QFC to telecom wavelengths.

\begin{figure}
\includegraphics[width=1\linewidth]{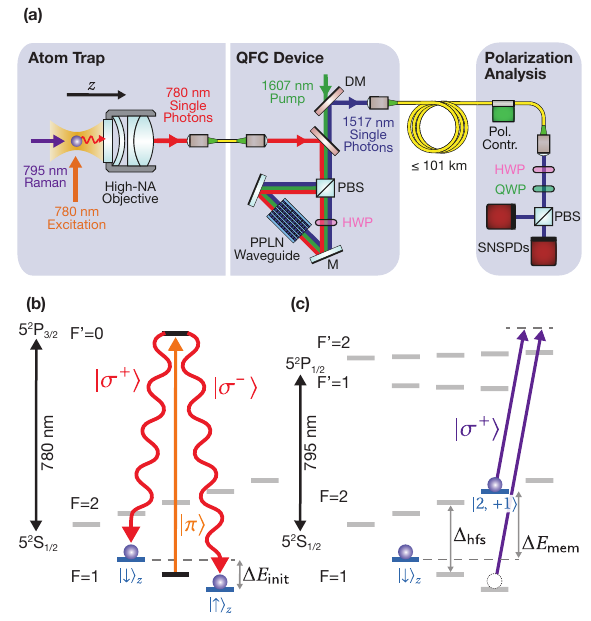}
\caption{\protect\textbf{Architecture of a neutral atom based quantum network node.}
(a) A single $^{87}$Rb atom is stored in a strongly focused optical dipole trap, and its fluorescence is collected by a high-NA objective (NA=0.5).
The photons emitted by the atom at \SI{780}{\nano\meter} are guided to a QFC device where they are converted to \SI{1517}{\nano\meter} while preserving the polarization state.
Next, the telecom photons propagate through up to \SI{101}{\kilo\meter} spooled fiber, after which their polarization is analyzed with a polarizing beam splitter (PBS) and superconducting nanowire single photon detectors (SNSPDs).
(b) Scheme of the atom-photon entanglement generation. A $\pi$-polarized laser pulse excites the atom, followed by a spontaneous decay to the $\ket{\downarrow}_z$ or $\ket{\uparrow}_z$ state. Here $\{\ket{\uparrow}_z,\ket{\downarrow}_z\}$ is defined as the initial basis for atomic qubit.
(c) Scheme of the Raman transfer for an atomic qubit. A Raman $\pi$-pulse, which couples $\ket{\uparrow}_z$ to $5^2S_{1/2}$ $\ket{F=2, m_{F} =+1}$ via a two-photon Raman process, changes the initial basis to the memory basis $\{\ket{\downarrow}_z, \ket{F=2, m_{F}=+1}\}$.
}
\label{fig:Fig1}
\end{figure}

\section{Quantum Network Node}
Our quantum network node consists of a single $^{87}$Rb atom trapped in a tightly focused optical dipole trap (ODT), loaded from a cloud of $^{87}$Rb atoms in a magneto-optical trap (Fig.~\ref{fig:Fig1} (a)).
The atom is prepared in the ground state $5^2$S$_{1/2}\ket{F=1, m_{F}=0}$ by optical pumping, from where it is excited to the state $5^2$P$_{3/2}\ket{F’=0, m_{F’}=0}$ (Fig. \ref{fig:Fig1} (b)). Subsequently, the atom spontaneously decays to the $F = 1$ manifold via different decay channels emitting a \SI{780}{\nano\meter} photon.
This photon, when collected along the quantization axis, is thus maximally entangled with the atom in the state
\begin{equation}
\ket{\Psi}_{\mathsf{Atom-Photon}} = \frac{1}{\sqrt{2}} \left(\ket{\downarrow }_z \ket{L}+\ket{\uparrow }_z \ket{R}\right),
\end{equation}

\noindent where $\ket{L}$ and $\ket{R}$ denote left- ($\sigma^+$) and right-circular ($\sigma^-$) photonic polarization states, respectively, and $\ket{\downarrow}_z=\ket{F=1, m_{F}=-1}$ and $\ket{\uparrow}_z=\ket{F=1, m_{F}=+1}$ denote the initial qubit basis \cite{Volz:06}. For more details about the experimental setup see Appendix~\ref{App:atom-trap}.

\subsection{Long-Lived Quantum Memory}
The coherence time of the atomic state encoded in this initial basis is strongly limited by two effects. 
The first is the position-dependent dephasing caused by an AC-Stark shift originating from longitudinal electric field components induced by the tightly focused ODT beam ($w_0=\SI{2.05}{\micro\metre}$) \cite{Wenjamin2008thesis,Thompson:13}. The second one is due to residual magnetic field fluctuations ($\sim$ \SI{0.5}{\milli\gauss}) around the atom in the lab environment, which result in Zeeman shifts causing additional dephasing. 
Although these effects only cause dephasing and thus do not destroy the coherence of the state right away, averaging over many runs in the experiment results in the same degradation of the fidelity of the state.
All this leads to typical coherence times of hundreds of microseconds~\cite{rosenfeld:11}.

To mitigate the first effect, the ODT intensity and thus its trap depth $\textit{U}$ is adiabatically lowered from 2.3 to \SI{0.19}{\milli\kelvin} before performing the excitation attempts, thereby reducing the longitudinal field components which scale linearly with the ODT intensity \cite{Kuhr:05}. The occupation probabilities of the vibrational levels are preserved if the trap depth is lowered adiabatically, therefore the temperature should be reduced (proportional to $\sqrt{U}$) \cite{Tuchendler:08}.
We confirmed this by measuring the temperature of the atom via the release and recapture method and observed a temperature decrease from \SI{30}{\micro\kelvin} to \SI{8}{\micro\kelvin} \cite{Tuchendler:08}.
The reduction in temperature also further reduces the effect of position-dependent dephasing especially concerning the transverse movement in the trap.
The second effect is counteracted by applying a bias field along the $z$-axis (optimized value of $B_z= \SI{244.5}{\milli\gauss}$) suppressing the magnetic field fluctuations perpendicular to it and, in addition, diminishing the residual longitudinal field components.
These measures result in a coherence time of $322.5\pm\SI{38.5}{\micro\second}$.

This coherence time is still not sufficient for achieving high fidelity over really long distances. 
To further increase the coherence time, we perform a transfer of the atomic qubit from the initial basis to a so-called memory basis defined by the states $\{\ket{\downarrow}_z, \ket{F=2, m_{F}=+1}\}$ \cite{Koerber:18}, as illustrated in Fig. ~\ref{fig:Fig1} (c).
In the small field regime, this reduces the magnetic field sensitivity of the atomic state by a factor $\chi \approx -  g_F/g_I = 503$, which for the applied bias field becomes $\chi= 545.6$, see Supplemental Material.
The qubit basis transfer is achieved by applying a pair of $\sigma^+$-polarized Raman beams coherently driving the transition between the atomic states $\ket{\uparrow}_z$ and $\ket{F=2, m_{F} =+1}$. For a single-photon detuning of $2\pi\times \SI{2.7}{\giga\hertz}$, we obtain an optimal $\pi$-pulse duration of \SI{8}{\micro\second}.
The magnetic bias field makes the transfer Zeeman-state selective due to the introduced energy-level splitting, see Appendix~\ref{App:raman}. 
The contrast of the Raman transfer, defined as the difference between residual populations after a single transfer pulse when initially prepared in $\ket{\downarrow }_z$ or $\ket{\uparrow }_z$, respectively, is measured to be 97.8$\pm$2.2\%. After a variable storage time, a second Raman pulse transfers the qubit state back to the initial basis.
For this process, it is essential that the two transfer processes are coherent, thus requiring phase stability of the difference frequency of the Raman pair during the storage time, 
which is achieved by using a low phase-noise microwave source referenced to a Rubidium atomic clock (see Appendix~\ref{App:Raman set-up}).
Finally, the atomic state is read out in the initial basis with a state-selective ionization scheme~\cite{Leent:22}.

Characterizing the coherence time of the atomic qubit in the memory basis, we find a coherence time of $6.91\pm \SI{0.42}{\milli\second}$, as shown in Fig.~\ref{fig:coherence} and further detailed in Appendix~\ref{App:coherence}.
The increase in coherence time is less than expected from magnetic field fluctuations alone, indicating other limiting sources of decoherence. Nonetheless, the significantly longer coherence time already allows entanglement distribution over fiber links with a length of almost a thousand kilometers.

\subsection{Telecom interface}
For bridging long distances, it is crucial to minimize attenuation losses in long fiber links. We thus convert the emitted 780 nm single photons to the low-loss telecom S-band, by employing difference frequency generation (DFG) where the field of a single photon is combined with a strong \SI{1607}{\nano\meter} pump laser field in a periodically poled lithium niobate (PPLN) waveguide obtaining a wavelength of \SI{1517}{\nano\meter} for the converted photon \cite{Leent:20,Leent:22}. To preserve the polarization of the single photons, the waveguide is placed in a Sagnac configuration, where $\ket{H}$ and $\ket{V}$ polarizations are converted independently but coherently without requiring additional stabilization. Anti-Stokes Raman scattering, induced by the pump laser in the waveguide, contributes wide-band background, which is, however, already strongly reduced at the pump--signal wavelength difference of \SI{-372.0}{\per\centi\meter}.
A spectral filter setup, including a Fabry-Perot cavity with a bandwidth of \SI{27}{\mega\hertz} FWHM, reduces the QFC induced background to 82.8 cps after a \SI{50}{\meter} fiber.
Altogether, an external device conversion efficiency of up to \SI{57}{\percent} is achieved~\cite{Leent:20}.

At last, the telecom photons are injected into a single-mode fiber on spools with a 0.2 dB/km loss (SMF-28 Ultra) and detected by a polarization analysis setup that contains superconducting nanowire single photon detectors (SNSPDs) with efficiencies $\eta>\SI{60}{\percent}$ and dark count rates of $R_{\mathsf{dc}}<\SI{6}{\cps}$. Polarization drifts of the single photons in the long fiber are actively compensated by using laser light at the single-photon frequency, a fiber polarization controller, and a polarimeter together with a gradient descent optimization algorithm~\cite{Leent:22}.

\begin{figure}
\centering
\includegraphics[width=1\linewidth]{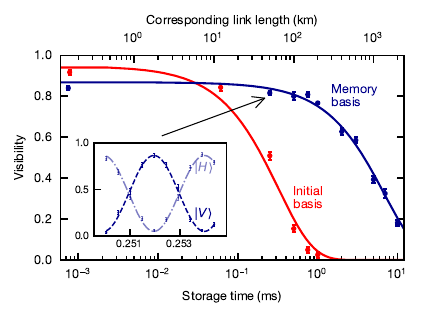}
\caption{\textbf{Characterization of the quantum memory coherence time.} Measured atom-photon state visibility as a function of the storage time. The red dots are measured visibilities in the initial basis $\{\ket{\uparrow}_z,\ket{\downarrow}_z\}$, achieved with ODT ramping and magnetic bias field. To obtain a longer coherence time, the atomic qubit is mapped into the memory basis via a Raman transfer (blue dots). The data are fitted with exponential decay functions showing an increase in coherence time from $322.5\pm\SI{38.5}{\micro\second}$ in the initial basis to $6.91\pm$\SI{0.42}{\milli\second} in the memory basis. The top axis indicates the corresponding fiber length by approximating the speed of the photon in the optical fiber with $\frac{2}{3}c$, where $c$ is the speed of light. The inset shows the atom-photon correlation for storage time around \SI{250}{\micro\second}.}
\label{fig:coherence}
\end{figure}

\begin{figure*}
\includegraphics[width=1\linewidth]{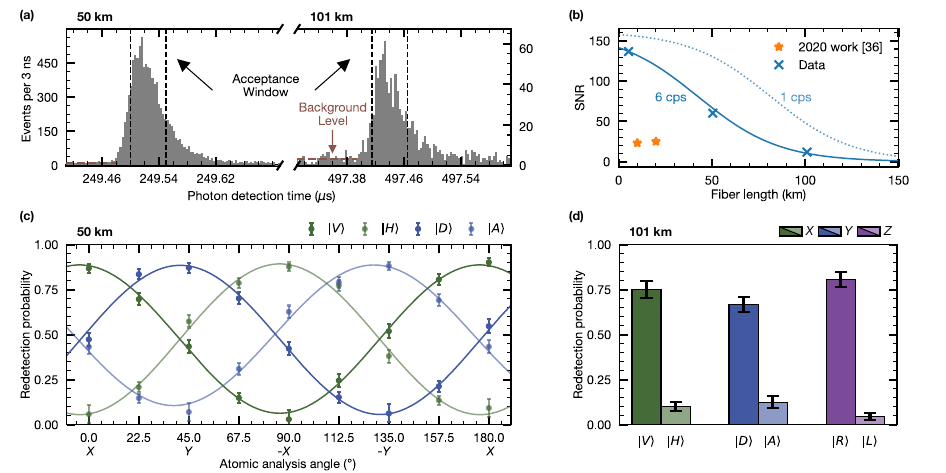}
\caption{\textbf{Quantum network node performance for entanglement distribution over 50 and \SI{101}{\kilo\meter} optical fiber.}  (a) Detection time histogram of the converted photons. Photon detections are accepted within a window of \SI{50}{\nano\second}, indicated with dashed black lines, accepting approximately \SI{62}{\percent} of the single photon events.
(b) SNR characterization as a function of fiber length. The lines represent a model of the SNR as a function of fiber lengths for $R_{\mathsf{dc}}=6$ and 1 cps. For completeness, we also measured the SNR for a fiber length of \SI{5}{\kilo\meter}. Compared to previous work \cite{Leent:20}, the QFC background is reduced by a factor of 8 due to a more favorable signal-pump wavelength combination. 
(c) Atom-photon state correlations over a fiber length of \SI{50}{\kilo\meter}. 
In the \SI{50}{\nano\second} photon window, 6548 photon detection events were recorded in 25.2 hours. The resulting estimated state fidelity is $\geq$84.8$\pm$\SI{1.1}{\percent}.
(d) Atom-photon state correlations over a fiber length of \SI{101}{\kilo\meter}. We recorded 656 photon detection events in 47.7 hours. The atom-photon correlations were measured in three bases, resulting in an estimated state fidelity of $\geq$70.8$\pm$\SI{2.4}{\percent}.
All error bars indicate one standard deviation.}
\label{fig:fibers}
\end{figure*}

\section{Long-Distance Entanglement Distribution}
The performance of the quantum network node was evaluated by analyzing the atom-photon entanglement distribution over fiber links of 50 and \SI{101}{\kilo\meter}. Fig.~\ref{fig:fibers}(a) shows the detection time histograms of the converted photons with arrival times of \SI{249.5}{\micro\second} and \SI{497.4}{\micro\second} after the entanglement generation, respectively. 
Detection events consequently trigger the atomic state readout without substantial delay.

If that detection event was due to the emitted photon during entanglement generation, one can observe a correlation with the measurement result of the atom. 
However, there are also erroneous detection events which consequently reduce the observed correlations and entanglement fidelity. 
Effectively, there are two noise sources, namely, the detector dark counts and the QFC background, of which the latter --- like the signal photons --- is attenuated in the fiber links.
To optimize the signal-to-noise ratio (SNR) (see Fig. \ref{fig:fibers}(b)) while still observing a sufficient event rate, we applied a \SI{50}{\nano\second} photon acceptance window in the data evaluation, accepting approximately 62\% of the detection events.
This resulted in an observed SNR of 60.3 and 11.8 for fiber lengths of 50 and \SI{101}{\kilo\meter}, respectively. Since the detector dark counts become the main noise source for long distances, installing lower-noise detectors would enable a better SNR. For example, as shown in Fig.~\ref{fig:fibers}(b), at $R_{\mathsf{dc}}=\SI{1}{\cps}$ we expect a SNR of 46.7 for 101 km fiber length. 

In the configuration with \SI{50}{\kilo\meter} fiber, $6548$ events were recorded. Given the additional attenuation of about a factor 10 --- the success probability was reduced to \SI{1.09e-4}, which, considering the repetition rate of \SI{1182}{\hertz}, resulted in an event rate of 1/\SI{14}{\per\second}. The entanglement fidelity relative to the state $\ket{\Psi}$ has been estimated by measuring correlation probabilities in two bases. For each measured polarization state of the photon (vertical $\ket{V}$, horizontal $\ket{H}$, and 45° $\ket{D}$, 135° $\ket{A}$ linear polarization, respectively), the atomic analysis angle has been rotated from the atomic $X$ basis via $Y$, $-X$ and $-Y$ back to $X$ in steps of \SI{22.5}{\degree} (Fig. \ref{fig:fibers} (c)). The data are fitted with sinusoidal curves yielding visibilities of 82.4$\pm$\SI{1.9}{\percent}, 83.7$\pm$\SI{2.8}{\percent}, 82.8$\pm$\SI{2.8}{\percent}, and 78.1$\pm$\SI{2.9}{\percent} for the $\ket{V}$, $\ket{H}$, $\ket{D}$, and $\ket{A}$ photon polarization states, respectively. This results in an averaged visibility of $\overline V = 81.8\pm1.3\%$, and a lower bound for the entanglement fidelity is estimated via $\mathcal{F} \geq \frac{1}{6} + \frac{5}{6} \overline V = 84.8\pm1.1\%$, relative to the maximally entangled state. The emergence of the factors $1/6$ and $5/6$ is due to the effective $2\times 3$ state space, as the $\ket{F=1, m_{F}=0}$ state orthogonal to $\ket{\Psi}$ can also be populated. In addition, these measurement settings also allow to evaluate the CHSH $S$ value \cite{Clauser:69}. With an $S=2.259\pm0.060$, the observed atom-photon state still violates the CHSH-Bell inequality with 4.3 standard deviations.

When distributing entanglement over a fiber length of \SI{101}{\kilo\meter}, the success probability of observing a photon emitted after an excitation pulse and transmitted over that distance decreased to \SI{1.08e-5}.. Due to the longer transmission time of the photon of \SI{497}{\micro\second}, the repetition rate had to be further decreased to \SI{910}{\hertz}, resulting in an event rate of only 1/\SI{262}{\per\second} (see Appendix~\ref{App:rate}). 
To estimate the entanglement fidelity, we examined the maximum correlation points by projecting the photonic polarization state in the eigenbasis \{$\ket{L},\ket{R}$\}, or one of the superposition bases \{$\ket{V},\ket{H}$\} and \{$\ket{D},\ket{A}$\}, while measuring the atomic qubit state in the respective orthogonal bases $X$, $Y$ or $Z$. For the correlators, we obtain $E_X=64.8\pm5.3\%$, $E_Y=54.2\pm5.3\%$, and $E_Z=76.0\pm4.5\%$, respectively. Together, these give a lower bound on the entanglement fidelity of $\mathcal{F} \geq \frac{1}{6} + \frac{5}{6} \overline V = 70.8\pm2.4\%$, where $\overline{V} = (E_X+E_Y+E_Z)/3$ is the average visibility. This result clearly exceeds the bound of $\mathcal{F}=50\%$ required to witness entanglement.

The decrease in the observed fidelity is attributed to the limited-SNR induced erroneous readout or uncorrelated detection events (6.6\%), decoherence of the atomic state (5.0\%), imperfect transfer efficiency of both Raman transfers (4.9\%), imperfect atomic state readout (4.7\%), entanglement generation errors (1.1\%), imperfect readout timing of the atomic state (0.8\%) and experimental drifts (4.0\%). Note that for the achieved coherence time, no substantial decrease in fidelity is observed over up to \SI{1}{\milli\second} (see Fig.~\ref{fig:coherence}), which is twice the photon travel time over \SI{101}{\kilo\meter}. This would be the required memory storage time for the distribution of heralded entanglement.

\section{Conclusion and Outlook}
This proof-of-concept experiment demonstrates a quantum network node that enables atom-photon entanglement distribution over fiber links up to 101 km.
Note the simultaneous submission by V. Krutyanskiy et al. ~\cite{Krutyanskiy2023}.
For the realization of efficient and large-scale quantum networks, however, the employed nodes should additionally allow to implement quantum repeater~\cite{Loock2020} and entanglement distillation protocols~\cite{reichle2006experimental, kalb2017entanglement}, which, require a substantially higher entanglement generation rate. 
The entanglement generation rate can be increased by improving the efficiencies of photon collection, transmission, and detection, e.g., by implementing an optical cavity~\cite{reiserer2015cavity} and more efficient single photon detectors. However, due to the latency of receiving information of the successful photon detection, this cannot increase the rate arbitrarily, which makes multiplexing of the entanglement generation process inevitable. For the neutral atom platform, a promising approach is to scale up the number of atoms via the creation of (multidimensional) atom arrays~\cite{madjarov2020high,bluvstein2022quantum}. Together with Rydberg~\cite{saffman2010quantum, huie2021multiplexed} or cavity enhanced interactions~\cite{ramette2022any} between the atoms, this will allow us to realize efficient large-scale quantum networks.

In this experiment, the coherence time of an atomic qubit could be increased to a level well suited for very long transmission distances by employing a state-selective Raman transfer to a basis much less sensitive to magnetic fields. Together with efficient polarization-preserving QFC, entanglement was observed between a single atom and a single photon that traveled through up to 101 km telecom fiber, thereby paving the way towards large-scale quantum networks based on heralded entanglement between distant single-atom quantum network nodes.

\section*{Acknowledgements}
We acknowledge funding by the German Federal Ministry of Education and Research (Bundesministerium f{\"u}r Bildung und Forschung (BMBF)) within the projects Q.Link.X (Contracts No. 16KIS0864, 16KIS0880), Q.R.X (Contracts No. 16KISQ001K and 16KISQ002), and QuKuk (Contracts No. 16KIS1621), by the Deutsche Forschungsgemeinschaft (DFG, German Research Foundation) under Germany’s Excellence Strategy – EXC-2111 – 390814868, and by the International Max Planck Research School (IMPRS).

\appendix
\section{SETUP AND SEQUENCE}
\label{App:setup}

\subsection{Single-atom trap}
\label{App:atom-trap}
Our single $^{87}$Rb atom is stored in an optical dipole trap. For this, a far-red-detuned light field ($\lambda$=\SI{849.5}{\nano\meter}) is focused down to a waist of \SI{2.05}{\micro\meter} by a custom-made high-NA objective (NA=0.5) which generates a trapping potential up to \SI{2.3}{\milli\kelvin} with intensity of \SI{4.97}{\milli\watt\per\micro\meter\squared}.
The trap is loaded from a cloud of atoms generated by a magneto-optical trap (MOT) at the focus position. The objective collects photons emitted by the atom in a confocal arrangement for entanglement distribution and atomic state detection. A micro-electromechanical systems (MEMS) fiber-switch directs the atomic fluorescence to a silicon avalanche photodiode (APD) during atom loading and readout, while the single photons are directed to the quantum frequency converter during the entanglement generation attempts.
For the state analysis the objective is also used for focusing the ionization beam. The ionization of the atom depends on the polarization of the readout beam. For more details about the state selective readout process, refer to \cite{Leent:22}.

\subsection{Raman state transfer setup}
\label{App:Raman set-up}
A pair of \SI{795}{nm} laser beam coherently drives the Raman transitions between two hyperfine states with a frequency difference of $\approx$ \SI{6.8}{\giga\hertz}. The beam pair is generated using a fiber-coupled intensity electro-optical modulator (NIR-MX800-LN-10, iXblue), which is temperature stabilized better than \SI{0.1}{\kelvin} and is driven by a microwave generator (SMA100B, Rohde \& Schwarz) referenced to a Rubidium atomic clock (E10-Y, Quartzlock).
A bias-voltage controller (MBC-DG-LAB-B2, iXblue) locks the operating mode of the EOM to achieve a minimal carrier intensity. The power of the Raman beam pair is actively stabilized using a PID (STEMlab 125-14, Redpitaya), with residual power drift of less than \SI{1.4}{\percent}.

\subsection{Experimental sequence}
\label{App:Sequence}
Once the single-atom trap is loaded, the atom-photon entanglement generation and analysis process starts. This process can be divided into three stages: (i) cooling, (ii) entanglement generation, and (iii) atomic state readout.
The cooling stage starts with \SI{1}{\milli\second} of polarization gradient cooling (PGC) at close to zero magnetic fields. Thereafter, the intensity of the cooling beams is ramped down in \SI{1.5}{\milli\second}, and the trap depth is adiabatically lowered from 2.3 to \SI{0.19}{\milli\kelvin} achieving atomic temperatures of around \SI{8}{\micro\kelvin}. Next, a magnetic bias field of \SI{244.5}{\milli\gauss} is set along the quantization axis which takes \SI{4}{\milli\second} to stabilize.
In the entanglement generation stage, the atoms are optically pumped to the ground state $5^2$S$_{1/2}\ket{F=1, m_{F}=0}$ with \SI{80}{\percent} efficiency in \SI{3}{\micro\second}, and then excited to the state $5^2$P$_{3/2}\ket{F’=0, m_{F’}=0}$ by a short laser pulse (Gaussian pulse, \SI{21}{\nano\second} FWHM) with an efficiency of \SI{90}{\percent}.
In the subsequent spontaneous decay (lifetime of the excited state \SI{26.24}{\nano\second}), the spin state of the single atom and the polarization state of the emitted single photon are entangled.
After each entanglement generation attempt, a \SI{8}{\micro\second} $\pi$-pulse of the Raman beams transfers the qubit encoding to the memory basis.
The quantum memory stores the quantum state with high fidelity while the photons travel through the long fibers, e.g., for a fiber length of \SI{100}{\kilo\meter}, transmission takes around \SI{500}{\micro\second}.
These cycles are repeated until a telecom photon is detected, which initiates the  (iii) atomic state readout. The atomic readout starts with a second Raman transfer that maps the qubit to the initial encoding followed by a state-selective readout scheme. After 11 unsuccessful attempts, the trap depth is increased to the initial depth and the magnetic field is set to close to zero in \SI{0.5}{\milli\second}, after which the cooling stage is restarted.

\begin{figure}
\centering
\includegraphics[width=1\linewidth]{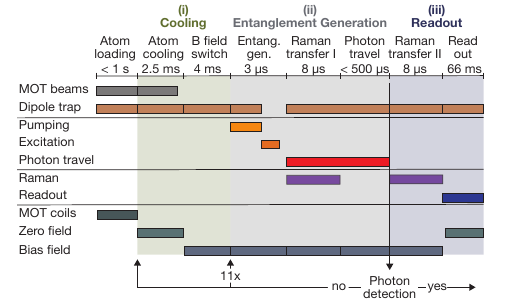}
\caption{\textbf{Experimental sequence to generate and analyze atom-photon entanglement.} The whole sequence is divided in three stages for cooling (green background), entanglement generation (light-gray background), and atomic readout (dark-gray background). The entanglement generation attempts are executed in a burst of 11, after which the atom is re-cooled. Successful detection of a single photon interrupts the burst and starts the atomic state readout. Raman transfer I coherently transfers the qubit from the initial to the memory encoding, while Raman transfer II maps the qubit back to the initial encoding before state readout.}
\label{fig:SNR}
\end{figure}

\section{RAMAN STATE TRANSFER}
\label{App:raman}
As presented in the main text, an essential technique is the Zeeman-state-selective Raman transfer.
It enables the swapping of the atomic qubit between
the  initial basis and the memory basis. The initial basis is used for entanglement generation and atomic state readout whereas the memory basis being less sensitive to magnetic field fluctuations is used for storing the quantum state. Hence, the memory basis encoding prolongs the atomic coherence time.

\begin{figure}
    \centering
    \includegraphics[width=1\linewidth]{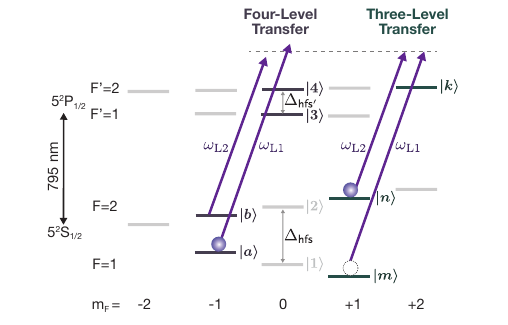}
    \caption{\protect\textbf{Energy level diagram of the Raman transfer scheme.} We utilize a pair of $\sigma^+$-polarized Raman beams on the $^{87}\text{Rb}$ $D_1$-line at \SI{795}{\nano\meter} wavelength to employ the Raman transfer, thus two transfers are possible.
For a three-level Raman transfer, we define $\ket{m}=\ket{F=1,m_F=+1}$, $\ket{n}=\ket{F=2,m_F=+1}$, and $\ket{k}=\ket{F'=2,m_{F'}=+2}$.
For a four-level Raman transfer, we define $\ket{a}=\ket{F=1,m_F=-1}$, $\ket{b}=\ket{F=2,m_F=-1}$, $\ket{3}=\ket{F'=1,m_{F'}=0}$ and $\ket{4}=\ket{F'=2,m_{F'}=0}$.
We also define $\ket{1}=\ket{F=1,m_F=0}$, $\ket{2}=\ket{F=2,m_F=0}$.
The energy difference between the two levels $\left|1\right\rangle$ and $\left|2\right\rangle$ ($\left|3\right\rangle$ and $\left|4\right\rangle$) is defined as $\Delta_\mathsf{hfs}$ ($\Delta_\mathsf{hfs'}$). Assuming a weak bias field $B_z$, the ground (excited) levels will be split with energy difference $g_F m_F\mu_B B_z$ ($g_{F'} m_{F'}\mu_B B_z$), where $|g_F|=1/2$ ($|g_{F'}|=1/6$).
The frequencies of the Raman pair are defined with $\omega_{\text{L1}}$ and $\omega_{\text{L2}}$, respectively.
}
    \label{fig:transfer-scheme}
\end{figure}

Fig. \ref{fig:transfer-scheme} shows a schematic of the Raman transfer scheme for $\sigma^+$ polarization. There, two different Raman transfer processes are possible, namely, the three-level or the four-level transfer.
The Hamiltonian for these processes in the rotating wave approximation is transformed into an effective two-level system (see Supplemental Material). There the following definitions for the detunings are employed: $\Delta_l=\omega_{Ll}-\omega_{ij}$ and $\overline{\Delta}=(\Delta_1+\Delta_2)/2$, where $\omega_{ij}=\omega_i-\omega_j$ with $l\in{1,2}$ and $i,j \in {m,n,k}$ or $\in {a,b,3,4}$ for the three- or four-level case, respectively.

The goal is to transfer only one of the two $m_F$-states to the corresponding $m_F$-state into the $F=2$ manifold. It turns out that this is best achieved using the three-level transfer by tuning the two-photon detuning $\delta$ for a given bias field $B_z$. 

\subsection{Three-level Raman transfer}
In the three-level transfer scheme, the atom defined in the levels $\left\{ \ket{m}, \ket{n}, \ket{k} \right\}$ interacts with two Raman beams $\left\{ \ket{\text{L1}}, \ket{\text{L2}} \right\}$.

Unity state transfer efficiency is only possible when the two diagonal items of the effective two-level Hamiltonian are equal, which results in
\begin{equation}
\delta_{\rm{3-level}} = 2g_F\mu_B B_z/\hbar + \dfrac{\left|\Omega_{nk}\right|^2-\left|\Omega_{mk}\right|^2}{4\overline{\Delta}-8g_{F'}\mu_B B_z/\hbar} ,
\label{delta:three-levels}
\end{equation}
where $\Omega_{ij}$ is the Rabi frequency for corresponding transition from state $\ket{i}$ to state $\ket{j}$, $\mu_B$ is the Bohr magneton and $g_F$ is the Landé g-factor.

\subsection{Four-level Raman transfer}
A $\sigma^+$-polarized Raman pair also enables the coherent population transfer from $\ket{a}$ to $\ket{b}$ via the two excited levels $\ket{3}$ and $\ket{4}$ in a four-level scheme.
Using the same method as above, the Hamiltonian of the four-level atom {$\left\{ \ket{a}, \ket{b}, \ket{3}, \ket{4}\right\}$} interacting with two Raman beams {$\left\{ \ket{\text{L1}}, \ket{\text{L2}} \right\}$} can be written as an effective two-level Hamiltonian.
Finally, to achieve optimal transfer efficiency in this case, the detuning needs to satisfy 
\begin{equation}
    \delta_{\rm{4-level}}= -2g_F\mu_B B_z/\hbar+\dfrac{\left|\Omega_{b3}\right|^2-\left|\Omega_{a3}\right|^2}{4\overline{\Delta}+4\Delta_\mathsf{hfs'}}+\dfrac{\left|\Omega_{b4}\right|^2-\left|\Omega_{a4}\right|^2}{4\overline{\Delta}}.
    \label{delta:four-levels}
\end{equation}

\subsection{Raman transfer implementation}
\begin{figure}
\includegraphics[width=1\linewidth]{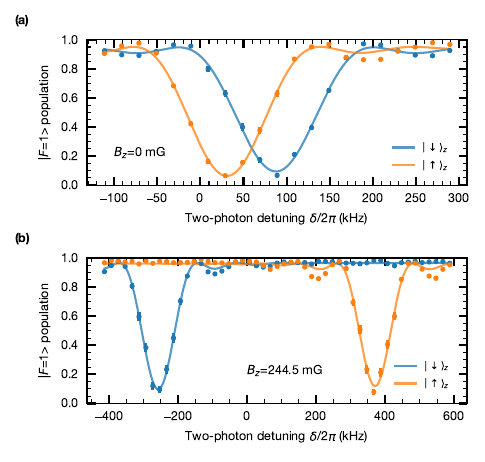}
\caption{\textbf{Characterization of the Zeeman-state-selective Raman transfer.}
Raman transfer for varying two-photon detuning $\delta$ at zero magnetic field (a) and with a bias field of $B_z=244.5$ mG (b). An atom is prepared in the $\ket{F=1}$ manifold in the state $\ket{\downarrow}_z$ (blue) or $\ket{\uparrow}_z$ (orange) state, after which the Raman transfer pulse is applied. A reduced $\ket{F=1}$ population signals a successful state transfer to the $\ket{F=2}$ manifold. During the experiments presented in the main text, a bias field is applied to achieve a Zeeman-state-selective transfer with a contrast of $97.8\pm2.2\%$.
The contrast is limited by the pulse power drift and polarization purity of the Raman transfer beam.}
\label{fig:Statetransfer}
\end{figure}

Eq.~\ref{delta:three-levels} and Eq.~\ref{delta:four-levels} show that for a given bias-field $B_z$ and average single-photon detuning $\overline{\Delta}$, both Raman transfers are optimized independently for two different two-photon detunings $\delta$.
We compared Raman transfer with far-detuned Raman light ($\overline{\Delta}$ $\approx$ \SI{2.7}{\giga\hertz}) with and without a bias field (Fig.~\ref{fig:Statetransfer}).
or no bias field, where the optimal detuning for the transfer of either $\ket{\uparrow}_z$ or $\ket{\downarrow}_z$ differs by the differential light shift of the $\sigma^+$-polarized Raman pair, the state-selectivity is limited (Fig.~\ref{fig:Statetransfer} (a)).
A clear separation is obtained for a bias field of \SI{244.5}{\milli\gauss} with a clear separation of the two-photon detuning frequencies (Fig.~\ref{fig:Statetransfer} (b)).
This configuration guarantees that by choosing the correct two-photon detuning $\delta$ one of the two states is transferred while the other one is blocked.

\section{COHERENCE TIME CHARACTERIZATION}
\label{App:coherence}

\begin{figure*}
\includegraphics[width=1\linewidth]{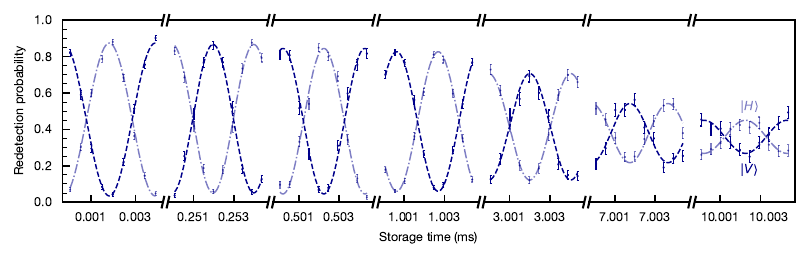}
\caption{\textbf{Characterization of the quantum memory coherence time in the memory basis.} We measure atom-photon correlations as a function of the storage time. The spin of the atomic state is read out in the $X$ basis, and the polarization of the photonic state is projected in the \{$\ket{V},\ket{H}$\} basis. We extract the visibilities by fitting the time evolution of the $\ket{V}$ state (dark blue) and the $\ket{H}$ state (light blue) with sinusoidal functions.
}
\label{fig:readoutdelay}
\end{figure*}

To characterize the coherence time of the atomic qubit in the initial and memory basis, we measure the temporal evolution of the atomic state in a bias field of \SI{244.5}{\milli\gauss} along the quantization axis at various delays, as shown in Fig. \ref{fig:readoutdelay} for the memory basis.
The experimental sequence starts by generating an atom-photon entangled state.
By projecting the photonic state on the \{$\ket{V},\ket{H}$\} basis, the atomic spin is prepared in the $X$-basis, i.e. in a superposition of the eigenstates in the magnetic field basis aligned along $z$.
Directly after the emission, the atomic state is transferred to the memory basis and then allowed to evolve freely over a programmable delay time.
Finally, the atomic qubit is coherently transferred back to the initial basis before being read in the $X$ basis.
The atomic state experiences Larmor precession due to the applied bias field which results in a time-dependent redetection probability in the $X$ basis.
For each delay, the visibility is obtained by fitting the data with sinusoidal curves.

Our atomic state detection scheme allows us to determine the atomic state in any basis, however we performed the measurements only in atomic $X$ basis, which is not the eigenbasis of the system. Therefore, we expect similar, if not better, visibilities in the other two bases. The obtained visibilities are plotted in Fig.~\ref{fig:coherence} and fitted with an exponential decay to obtain the $T_2$ time of the state.

\section{ENTANGLEMENT GENERATION RATE} 
\label{App:rate}

The atom-photon entanglement generation rate $r$ is given by
\begin{equation}
r=\phi\cdot R\cdot\eta,\\
\end{equation}
where $\phi$ is the duty cycle of the experiment, $R$ is the repetition rate of the entanglement generation attempts, and $\eta$ denotes the success probability of detecting a photon for each entanglement generation try.

The repetition rate of the entanglement generation attempts is limited by
\begin{equation}
R_{\rm{max}}(L)=\frac{1}{T(L)}=\frac{1}{T_{L=0}+L/c_f},\\
\end{equation}
where $T(L)$ is the duration of an entanglement generation try, $L$ is the fiber length, $c_f$ is the speed of light in the optical fiber, which is in good approximation about $\frac{2}{3}c\approx \SI{200000}{\kilo\meter\per\second}$.
Here, $T_{L=0}$ is the period of an entanglement generation try for a link with $L=0$. It includes the preparation of the initial qubit state (\SI{3}{\micro\second}), the generation of the entanglement (\SI{200}{\nano\second}), the duration of Raman transfer (\SI{8}{\micro\second}), and the time to cool the atom per excitation try (\SI{6.5}{\milli\second}/11) resulting in a $T_{L=0}=$\SI{602.12}{\micro\second}.
In this work, we thus used repetition rates of \SI{1.59}, \SI{1.18}, \SI{0.91}{\kilo\hertz} for \SI{5}, \SI{50}, \SI{101}{\kilo\meter} fiber lengths, which are limited by the travel time of the photon through the optical fiber, and the switching time of the magnetic field from zero field to magnetic guiding field (\SI{4}{\milli\second}) which can be optimized independently for future works. For more details see Appendix~\ref{App:Sequence}. 

\begin{table}
\centering
\caption{\textbf{Observed entanglement generation success probabilities and rates.} Listed are the observed success probabilities to generate atom-photon entanglement ($\eta$) and the entanglement generation rate ($r$), for the different fiber link lengths $L$. The success probabilities contain only photon detection events in the \SI{50}{\nano\second} acceptance window.}
\bgroup
\def\arraystretch{1.75}%
\begin{center}
\begin{tabular}{ c c c } \hline
L (km) & $\eta$                  & $r$ (s$^{-1}$)              \\ \hline
5      & $0.8277 \times 10^{-3}$ & 1/3    \\
50     & $0.109 \times 10^{-3}$  & 1/14   \\
101    & $0.0108 \times 10^{-3}$ & 1/262 \\ \hline
\end{tabular}
\end{center}
\egroup
\label{tab:prob}
\end{table}

The success probability to generate a single photon at \SI{1517}{\nano\meter} for each entanglement generation try equals
\begin{equation}
\eta(L)=\eta_{L=0}\cdot e^{-\frac{\alpha}{10}L},\\
\end{equation}
where $\eta_{L=0}$ is the success probability with $L=0$ and $\alpha$ equals the attenuation factor in the telecom fiber. Here, $\eta_{L=0}$ includes 
the collection and coupling efficiencies of the single photons emitted by the single atoms per excitation attempt (\SI{1.0}{\percent}), the transmission of the MEMS switch (\SI{85}{\percent}), the total efficiency of the QFC setup (\SI{48}{\percent}, not optimally aligned), the transmission efficiency of the single photon through a filter cavity (\SI{82}{\percent}), the transmission efficiency of the photon projection setup and fiber coupling (\SI{85}{\percent}), the three fiber-fiber connectors (\SI{94}{\percent}) and the average detection efficiency of the SNSPDs (\SI{59.7}{\percent}). Using QFC, the attenuation rate is reduced to \SI{0.2}{\decibel\per\kilo\meter} at \SI{1517}{\nano\meter}, such that we obtain a transmission of \SI{1.0}{\percent} when photons travel through the \SI{101}{\kilo\meter} fiber. 
The obtained success probabilities for the different fiber lengths within an acceptance photon window of \SI{50}{\nano\second} are listed in Tab.~\ref{tab:prob}.

The duty cycle is approximately 1/2 in the experiment, including the time required for the polarization compensation of the long fibers, the magnetic field stabilization, the Raman pulse power stabilization, and the fraction of time used to verify whether an atom is present in the trap or not.

In the full hard-wired acceptance window (\SI{208}{\nano\second}), we recorded \num{11939}, \num{9983}, and \num{1058} events within \num{5.8}, \num{25.2}, and \num{47.4} hours for \SI{5}, \SI{50}, \SI{101}{\kilo\meter} fiber link lengths, respectively, resulting in an event rate of 1/2~$s^{-1}$, 1/9~$s^{-1}$, 1/162~$s^{-1}$ for coincidence detection. The obtained event rates in the \SI{50}{\nano\second} acceptance window are listed in Tab.~\ref{tab:prob}.

\section{SIGNAL AND NOISE DETECTION EVENTS} 
\label{App:snr}
The entanglement observable between atom and photon is easily degraded by various noise sources. It is thus essential to analyze the probability of detecting a single photon, background originating in the QFC process, and the detector dark counts in the \SI{50}{\nano\second} photon window used in this experiment.
Like the single photons, background from the QFC propagates in the optical fiber and is thus equally attenuated (0.2 dB/km).
Therefore, in our setup, for fiber lengths significantly longer than \SI{50}{\kilo\meter}, the dark counts of the detectors which does not dependent on fiber length become the primary noise source, as shown in Fig.~\ref{fig:SNR}.
As mentioned in the main text, the next level of long distance entangelement distribution requires both further improvement of the photon collection but as well as very low noise single photon detectors without sacrificing detection efficiency.

\begin{figure}
\centering
\includegraphics[width=1\linewidth]{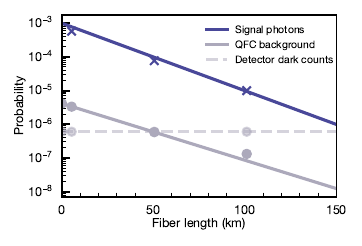}
\caption{\textbf{Detector click probabilities for varying fiber length.} Three processes can be distinguished that generate single-photon detector clicks: the signal photons (blue), noise introduced by the QFC device (gray solid line), and noise introduced by the detectors, i.e., detector dark counts (gray dashed line). In our setup, the detector dark counts become the prominent noise source for fiber lengths above \SI{50}{\kilo\meter}.}
\label{fig:SNR}
\end{figure}


%

\end{document}